\documentclass[12pt]{article}

\usepackage{graphicx}
\usepackage{newtxtext}
\usepackage{newtxmath}
\usepackage{hyperref}
\hypersetup{
colorlinks = true,
urlcolor   = blue,
citecolor  = black,
}

\newcommand{\RomanNumeralCaps}[1]
\linenumbers

\usepackage[margin=2.5cm]{geometry}
\usepackage{amsmath}
\usepackage{bm}
\usepackage{caption}
\usepackage{subcaption}
\usepackage{parskip}
\usepackage{verbatim}
\usepackage{xcolor}
\usepackage{mathtools}
\usepackage{setspace}

\definecolor{plotblue}{RGB}{0,122,153}
\definecolor{plotred}{RGB}{255,64,64}
\definecolor{plotgreen}{RGB}{51,178,102}

\title{Viscoelasticity characterization of compressible soft matter via fluid-mediated dynamic interactions}

\author{Pratyaksh Karan, Jeevanjyoti Chakraborty, Suman Chakraborty \\ \\ Department of Mechanical Engineering \\ Indian Institute of Technology Kharagpur \\ West Bengal (India)}

\date{}

\begin{document}
\maketitle

\onehalfspacing

\begin{abstract}
Characterizing the softness of deformable materials having partial elastic and partial viscous behaviour via soft lubrication experiments has emerged as a versatile and robust methodology in recent times. However, a straightforward extension of the classical elastohydrodynamic lubrication theory that is commonly employed for characterizing elastic materials turns out to be rather inadequate in explaining the response of such viscoelastic materials subjected to dynamic loading conditions, despite adhering to a mathematically acceptable framework via the complex Young's modulus as a material property. This deficit stems from a non-trivial interplay of the material compressibility and its time-dependent dynamic response under fluid-mediated oscillatory loading typical to surface probing experiments. Here we develop a soft-lubrication based theoretical framework that enables the consistent recovery of viscoelastic model properties of materials from experimental data, independent of the specific loading condition. A major advancement here is the rectification of inconsistencies in viscoelasticity characterization of previously reported models that are typically manifested by unphysical dependencies of the model parameters on the substrate layer thickness and the oscillation frequency of the surface probing apparatus. Our results provide further pointers towards the design of characterization experiments for consistent and specific mapping of the experimental data with the parametric values of the chosen material-model. These findings appear to be imperative in designing and selecting materials for emerging bio-engineering tools including organ-on-a-chip and human body-on-a-chip.
\end{abstract}

\section{Introduction}\label{sec:Introduction}

% Scanning probe microscopy (SPM) has emerged as a versatile class of techniques for probing objects and structures at the micro and nano scale. While the initial developments in SPM techniques focussed strongly on obtaining high resolution topographical data of miniature structures, these techniques eventually became diversified for a plethora of modalities of investigating small-scale systems. Examples include quantification of colloidal forces, studies on adhesion and rebound dynamics, rheometry of Newtonian and complex fluid and determination of constitute mechanical nature of soft materials. $[$about theoretical$]$

% Over the course of a little more than a decade, Leroy Charlaix and their group have developed a class of SPM techniques for probing soft coatings without direct contact, by transmitting 

Characterizing mechanical properties of soft materials is critical in various applications like drug delivery, flexible coating, biomimetic microdevices, to name a few \cite{Garcia2016,Liang2019,Liu2020,Saintyves2020,Chudak2020}. These materials are constitutively complex in their mechanical behaviour, in the sense that these cannot be comprehensively described by exclusive `solid-like' or `fluid-like' characteristics \cite{Guan2017,Guan2017a,Wintner2020,Kargar2021,Pandey2016}. Broadly termed as viscoelastic materials, their stress-responsive characteristics are intrinsically time-dependent, triggering a plethora of dynamic phenomena that are not observed for elastic solids. This necessitates specialized interpretation of experimental data for material characterization, as opposed to the standardized methodologies for elastic solids. Over and above, a generic theoretical depiction deems imperative to extract the resultant rheological parameters from the experimental observables, ironing out any possible anomaly stemming from inconsistent assumptions in the underlying conceptual premise.

Fluid-mediated surface probing apparatus, commonly utilized for characterizing soft materials without involving solid to solid adhesive contact and obviating the needs of delicate sample preparation, is progressively being developed to provide the essential foundations for determining the constitutive model properties of soft materials \cite{Butt2005,Leroy2011,Leroy2012,Kaveh2014,Carpentier2015,Guan2017a,Wang2017a,Tan2019,Mahadevan2005,Garcia2020}. Obtaining these material properties holds the key towards understanding several biologically relevant processes, including but not limited to cellular events, food ingestion, tearing in the eyes and functioning of the synovial joints \cite{vanAken2010,Dedinaite2012,Moyle2020,Zuk2021}. More recently, bio-compatible surfaces are being functionalized with soft coatings to create bio-engineered microenvironments for in-vitro analytics of microvascular physiology and cellular dynamics, in an effort to derive key insights on a plethora of diseased conditions ranging from arterial blockages to cancer metastasis \cite{Karan2020b,Moyle2020,Priyadarshani2021}.

Materials with pure elastic attributes are classically characterized in terms of a combination of relevant constitutive parameters, for instance, the combination of Young's modulus and Poisson's ratio \cite{Leroy2011,Cuddalorepatta2020}. These parameters remain constants for homogeneous and isotropic materials, irrespective of the nature of loading, over the linear regime of operation \cite{Salenccon2012}. Similar conceptual paradigms should ideally apply for the parameters characterizing viscoelastic materials as well. However, reported theories for viscoelastic material characterization based on contact-free experiments fail to capture this universality in the material parameters independent of the experiment conditions \cite{Guan2017,Guan2017a}. These discrepancies arise because the mentioned theories recover a simple Young’s modulus assuming the material to be incompressible, without delving into the effect of material compressibility and the association of the recovered complex Young’s modulus to the parameters of a chosen viscoelastic constitutive model.

Overcoming these limitations, here we develop a theoretical framework that addresses the soft lubrication problem for a physical setup identical to that used by Guan {\it et al} \cite{Guan2017}, considering the substrate material to be linearly-viscoelastic. By accommodating the substrate material compressibility into the mathematical model, we resolve the anomaly of dependence of the substrate material viscosity on the substrate layer thickness, as reported by Guan {\it et al} \cite{Guan2017}. On similar lines, we demonstrate that a possible anomaly of dependence of the substrate viscoelastic properties on the oscillation frequency of the probing apparatus can be resolved by using the standard linear solid model rather than the Kelvin-Voigt model to represent the viscoelastic constitution. Additionally, we pinpoint preferred experimental conditions for such material characterization. Towards this, we first show that by using a thin substrate layer, substrate material compressibility effects may be consistently accounted for. Second, by aptly-tuning the oscillation frequency, the precision in characterization of the relaxation time of the substrate material (pertinent to the standard linear solid model) gets improved. These inferences appear to be imperative in designing and advancing novel materials for bioengineering and healthcare research.

\section{Modeling}\label{sec:Math}

\subsection{Setup and Governing Equations}\label{subsec:model}

\begin{figure}[!htb]
\centerline{\includegraphics[width=0.8\linewidth]{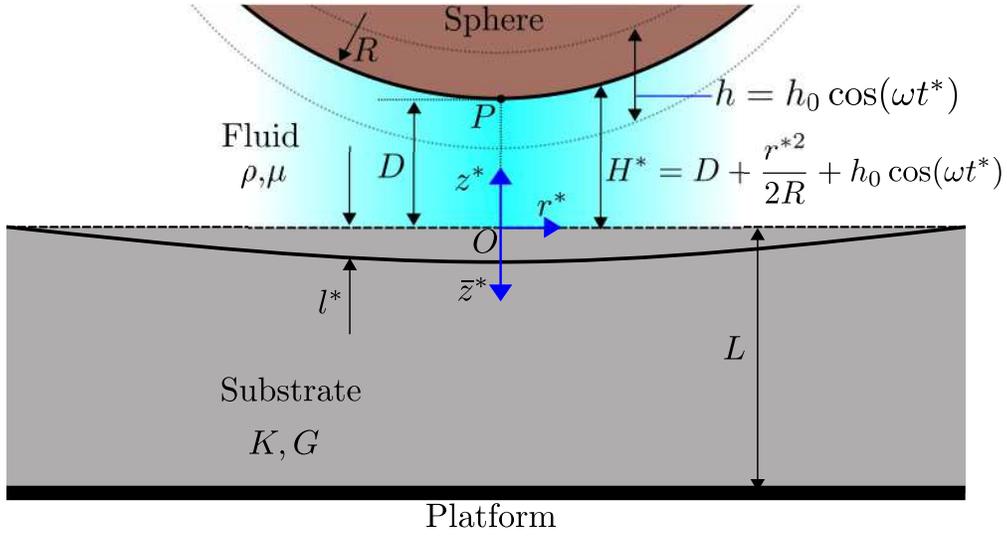}}
\caption{Schematic representation of the physical setup. The spherical probe (depicted as the brown sphere) of radius $R$ oscillates, with amplitude $h_0$ and frequency $\omega$, above the deformable substrate layer (depicted as the grey layer) having thickness $L$ in its un-deformed configuration, made of homogeneous, isotropic, compressible, viscoelastic material, mediated by a thin layer of viscous liquid. Constitutively, the substrate material properties are represented by the moduli functions ${K}$ and ${G}$, which are akin to the bulk modulus and shear modulus for linear-elastic materials. The intervening liquid is homogeneous isotropic, incompressible and Newtonian with density and viscosity as $\rho$ and $\mu$ respectively. $D$ represents mean separation of the sphere from origin.
% \protect\\ and (\textit{b}) half-periodic solutions.
}
\label{fig:schematic}
\end{figure}

The setup considered here is schematically presented in figure \ref{fig:schematic}. Time is denoted by $t^*$. We use the $r^*-z^*$ coordinate system for the fluid domain, and, the $r^*-\bar{z}^*$ co-ordinate system for the substrate domain. The relevant system variables are denoted as follows: $\vec{v}^*$ is the fluid velocity field, $\vec{u}^*$ is the substrate displacement field, $p^*$ is the pressure in the fluid domain, $\displaystyle F^* \left( = \int_{0}^{\infty}2\pi p^*r^*{\rm d}r^*\right)$ is the force between the sphere and the substrate, and $l^*$ is the fluid-substrate interface deflection, henceforth referred to as `deflection'. Three key ratios of system parameters are presented in the leftmost two columns of table \ref{tab:nondim}. Other relevant symbols are defined in the caption of figure \ref{fig:schematic}.

The sphere oscillation leads to an instantaneous gap height of $H^*$ between the sphere and the substrate \cite{Leroy2011} (see figure \ref{fig:schematic}),

\begin{equation}
\label{eq:H}
H^* = D+\frac{r^{*2}}{2R}+h_0\cos(\omega t^*) 
\end{equation}

The flow dynamics of the intervening fluid are governed by the continuity equation,
\begin{equation}
\label{eq:continuity_compact}
\nabla^*\cdot\vec{v}^* = 0,
\end{equation}
and Stokes equation,
\begin{equation}
\label{eq:momentum_compact}
0=-\nabla^* p^* + \mu\nabla^{*2}\vec{v}^*,
\end{equation}
for an incompressible Newtonian fluid, closed by the boundary conditions,
\begin{align}
\label{eq:conditions_compact}
v_r^* = 0, ~ v_z^* = -\omega h_0 \sin(\omega t^*) & \text{ \hspace{25pt} at \hspace{15pt} } z^*= H^*+l^*, \\
v_r^* = 0, ~ v_z^* = -\frac{\partial l^*}{\partial t^*} & \text{ \hspace{25pt} at \hspace{15pt} } z^* = -l^*, \\
v_r^* \rightarrow 0,~v_z^* \rightarrow 0,~p^* \rightarrow 0 & \text{ \hspace{25pt} for \hspace{15pt} } r^* \gg \sqrt{DR}, \\
v_r^* = \frac{\partial v_z^*}{\partial r^*} = \frac{\partial p^*}{\partial r^*} = 0 & \text{ \hspace{25pt} at \hspace{15pt} } r^* = 0.
\end{align}
The expanded forms of equations \eqref{eq:continuity_compact} and \eqref{eq:momentum_compact} are presented in section S1 of the ESI. 

The deformation behaviour of the substrate is governed by the mechanical equilibrium equation, 
\begin{equation}
\label{eq:mecheq}
\nabla^*\cdot\underline{\underline{\sigma}}_{S}^* = 0,
\end{equation}
where $\underline{\underline{\sigma}}_{S}^*$ is the substrate domain Cauchy-Green stress tensor, related to the strain tensor in the substrate layer $\displaystyle {\underline{\underline{E}}_{S}^*} = \frac{1}{2}\left(\nabla^* \vec{u}^*+(\nabla^* \vec{u}^*)^{\rm T}\right)$ as per the hereditary integral formulation\cite{Ferry1980},
\begin{equation}
\label{eq:heriditary}
\begin{split}
& \underline{\underline{\sigma}}_{S}^*(r^*,\bar{z}^*,t^*) = \\ \int_{-\infty}^{t^*} {\rm d}\tau^*   \bigg[ &  \left({K}(t^*-\tau^*)   - \frac{2}{3}{G}(t^*-\tau^*)\right)  \frac{\partial}{\partial\tau^*}\left\{\text{tr}\left({\underline{\underline{E}}_{S}^*(r^*,\bar{z}^*,\tau^*)}\right)\right\}\underline{\underline{I}} + \\ &~2{G}(t^*-\tau^*) \frac{\partial }{\partial \tau^*}\left\{{\underline{\underline{E}}_{S}^*(r^*,\bar{z}^*,\tau^*)}\right\} \bigg],
\end{split}
\end{equation}
where the superscript T implies matrix transpose, $\tau^*$ is the additional temporal variable used for retaining the deformation history of the material in the model considerations, and ${G}$ and ${K}$ are the viscoelastic shear and bulk modulus functions respectively for the substrate material. We emphasize that in this constitutive depiction, the current as well as historic state of strain determines the current state of stress, a key feature of viscoelastic materials. These equations are closed by the boundary conditions,
\begin{align}
\label{eq:solid_conditions_compact}
u_r^*\rightarrow 0,~u_{\bar{z}}^* \rightarrow 0 & \text{ \hspace{25pt} for \hspace{15pt} } r^* \gg \sqrt{DR}, \\
u_r^* = u_{\bar{z}}^* = 0 & \text{ \hspace{25pt} at \hspace{15pt} } \bar{z}^* = L, \\
u_r^* = \frac{\partial u_{\bar{z}}^*}{\partial r^*} = 0 & \text{ \hspace{25pt} at \hspace{15pt} } r^*=0, \\
\label{eq:solid_conditions_traction_compact} {\underline{\underline{\sigma}}_{S}^*}\cdot\hat{n} = {\underline{\underline{\sigma}}_{F}^*}\cdot\hat{n} & \text{ \hspace{25pt} at \hspace{15pt} } \bar{z}^* = 0,
\end{align}
where ${\sigma_{F}^*}$ is the fluid domain stress tensor. The expanded forms of equations \eqref{eq:mecheq} and \eqref{eq:solid_conditions_traction_compact} are presented in section S2 of the ESI. 

\subsection{Reynolds Equation and Pressure-Deflection Relation}\label{subsec:simp}

All the governing equations and boundary conditions are non-dimensionalized as per the characteristic scales of the variables, given in table \ref{tab:nondim}  for reference. The normalized governing equations and boundary conditions are presented in section S3 of the ESI. Subsequently, we consider certain simplifying assumptions which are standard for the setup being studied here. These assumptions are listed below.
\begin{enumerate}
\item We consider $\alpha \ll 1$ i.e., the oscillation amplitude is small compared to the sphere-substrate separation. This is typical to the experimental technique in consideration\cite{Guan2017}. 
\item We assume $\epsilon^{1/2} \ll 1$, i.e. the fluid dynamics in the gap between the sphere and substrate satisfies the lubrication approximation. This assumption has been reported to be adequate for the purpose of predicting the force responses irrespective of the instantaneous distances of separation between the confining solid boundaries.
\item We assume that the substrate deformation is negligible compared to the sphere-substrate separation, i.e. $\displaystyle \theta \ll \epsilon \implies \Gamma = \frac{\theta}{\epsilon} \ll 1$. This is in line with the consideration of Leroy \& Charlaix \cite{Leroy2011}.
\end{enumerate}

\begin{table}
\begin{center}
\def~{\hphantom{0}}
\begin{tabular}{cccccc}
\hline
\textbf{Ratio}					&	\textbf{Notation}									&
\textbf{Variable}				&	\textbf{Scale}										&	
\textbf{Variable}				&	\textbf{Scale}										\\ [3pt]
\hline
$\displaystyle D/R$   			& 	$\displaystyle \epsilon$							&
$z^*, H^*$						&	$\epsilon R$										&	
$\displaystyle \bar{z}^*$		& 	$\displaystyle \delta R$							\\
$\displaystyle h_0/D$   		& 	$\displaystyle \alpha$								&	
$v_r^*$							&	$\displaystyle \epsilon^{\frac{1}{2}}\alpha\omega R$&	
$\vec{u}^*, l^*$ 				& 	$\displaystyle \theta R$							\\
$\displaystyle L/R$   			& 	$\displaystyle \beta$ 								&	
$v_z^*$							&	$\displaystyle \epsilon\alpha\omega R$				&	
$r^*$							&	$\epsilon^{\frac{1}{2}}R$ 							\\
~								& 	~					 								&	
$p^*$							&	$\displaystyle \frac{\mu\omega\alpha}{\epsilon}$	&	
$t^*,~\tau^*$					&	$\displaystyle \frac{1}{\omega}$					\\
\hline
\end{tabular}
\caption{Assigned notations of pertinent ratios, and, characteristic scales of system variables; $\theta=\displaystyle \frac{\delta\alpha}{\epsilon}\frac{3\mu\omega}{(3{K}_c+4{G}_c)}$, $\displaystyle \Gamma = \frac{\theta}{\epsilon}$, and $\delta=\min(\beta,\epsilon^{\frac{1}{2}})$.}
\label{tab:nondim}
\end{center}
\end{table}

Following lubrication approximation, the simplified governing equations and boundary conditions yield the following governing equation for pressure distribution in the fluid domain (derivation present in section S4 of ESI),
\begin{equation}
\label{eq:Re_eq_simp}
\frac{1}{12r}\frac{\partial}{\partial r}\left\{r\left(1+\frac{r^2}{2}\right)^3 \frac{\partial p}{\partial r}\right\} = \frac{\Gamma}{\alpha}\frac{\partial l}{\partial t}-\sin(t),
\end{equation}
which is subject to the boundary conditions,
\begin{equation}
\label{eq:phd_bcs_main}
\left.\frac{\partial p}{\partial r}\right|_{r=0} = 0,~~~~\left.p\right|_{r\rightarrow \infty} \rightarrow 0.
\end{equation}

On the other hand, the pertinent governing equations and boundary conditions for substrate deformation are subjected to the Elastic-Viscoelastic Correspondence Principle (EVCP) \cite{Phan2017,Tschoegl2002}, yielding the frequency-domain equations. Next, these frequency-domain equations are subjected to Hankel-space solution of axisymmetric elasticity equations \cite{Zhao2009}. The derivation is presented in section S5 of ESI, and the final relation between fluid pressure and substrate deflection is,
\begin{equation}
\label{eq:l_H0}
\tilde{\hat{l}} = \breve{X}\tilde{\hat{p}}.
\end{equation}
Here, the accent $~\tilde{}~$ represents zeroeth-order Hankel transformation, 
\begin{equation}
\label{eq:Hankel_maindoc}
\tilde{\Psi}(x) = \int_{0}^{\infty}{\rm d}r~~ rJ_{0}(xr)~~\Psi(r),
\end{equation}
and the accent $~\hat{}~$ represents Fourier transformation, 
\begin{equation}
\label{eq:fouriertransform_maindoc}
\hat{\Phi}(f) = \int_{-\infty}^{\infty}{\rm d}t\exp(-2\pi ift)\Phi(t).
\end{equation}
In equation \eqref{eq:l_H0} above, $\displaystyle \breve{X}$ is the Fourier-Hankel space compliance function, and it represents the combined effect of substrate material viscoelasticity and substrate layer thickness. The expression for $\displaystyle \breve{X}$ is,
\begin{equation}
\label{eq:X}
\breve{X} =~~ \frac{\beta}{3\delta\bar{\mathcal{G}}}\frac{{\rm N}^{\underline{\rm r}}}{{\rm D}^{\underline{\rm r}}},
\end{equation}
with ${\rm N}^{\underline{\rm r}}$ and ${\rm D}^{\underline{\rm r}}$ given as,
\begin{subequations}
\begin{equation}
\label{eq:X_Nr}
{\rm N}^{\underline{\rm r}} = ~~ \Lambda_{(7,4)}\left(1-\exp\left(-4\bar{\beta}x\right)\right)-4\Lambda_{(1,4)}\bar{\beta}x \exp\left(-2\bar{\beta}x\right),
\end{equation}
\begin{equation}
\label{eq:X_Dr}
\begin{split}
{\rm D}^{\underline{\rm r}} = & ~~ 2\Lambda_{(7,1)}\bar{\beta}x\left(1+\exp\left(-4\bar{\beta}x\right)\right)+ \\
& \left\{4(\Lambda_{(4,4)}+9\bar{\mathcal{G}}^2)\bar{\beta} x+8\Lambda_{(1,1)}\bar{\beta}^3x^3\right\}\exp\left(-2\bar{\beta}x\right), \\
\end{split}
\end{equation}
\end{subequations}
where $\Lambda_{(m,n)} = (3\bar{\mathcal{K}}+m\bar{\mathcal{G}})(3\bar{\mathcal{K}}+n\bar{\mathcal{G}})$,
\begin{subequations}
\label{eq:coeffs}
\begin{equation}
\label{eq:coeffs_K}
\bar{\mathcal{K}}(f) = 2\pi if\int_{0}^{\infty}\frac{\exp(-2\pi ift){K}(t)}{3K_c+4G_c}~{\rm d}t,
\end{equation}
\begin{equation}
\label{eq:coeffs_G}
\bar{\mathcal{G}}(f) = 2\pi if\int_{0}^{\infty}\frac{\exp(-2\pi ift){G}(t)}{3K_c+4G_c}~{\rm d}t,
\end{equation}
\end{subequations}
$\displaystyle \bar{\beta} = \frac{\beta}{\sqrt{\epsilon}}$ and $K_c$ and $G_c$ are constants having the dimension of $K_0$ and $G_0$, and their expressions are presented ahead in section \ref{sec:moduli}. 

The $\bar{\mathcal{K}}$ and $\bar{\mathcal{G}}$ here explicate the role of the viscoelasticity of the substrate. These are discussed in \ref{sec:moduli}, and their pertinent implications in enabling coherent interpretations from the viscoelasticity characterization experiments are discussed in section \ref{sec:results}.

We solve the set of reduced governing equations in the Fourier space, by additionally subjecting equations \eqref{eq:Re_eq_simp} and \eqref{eq:phd_bcs_main} to Fourier transformation, yielding,
\begin{equation}
\label{eq:Re_eq_simp_Fourier}
\frac{1}{12r}\frac{\partial}{\partial r}\left\{r\left(1+\frac{r^2}{2}\right)^3 \frac{\partial \hat{p}}{\partial r}\right\} = 2\pi if \hat{l}-\text{Fr}[\sin(t)],
\end{equation}
\begin{equation}
\label{eq:phd_bcs_main_Fourier}
\left.\frac{\partial \hat{p}}{\partial r}\right|_{r=0} = 0,~~~~\left.\hat{p}\right|_{r\rightarrow \infty} \rightarrow 0.
\end{equation}

\subsection{Viscoelastic Moduli Functions}\label{sec:moduli}

We first appeal to the `Kelvin-Voigt' (KV) model as a representative constitutive framework in which the stress is considered to be a superposition of strain-dependent and rate-of-strain-dependent linearized functions. Boltzmann criticized the lack of generality in this approach, as it is constrained to quantify the stress in a material in terms of only the current strain and the strain at an infinitesimally previous moment. In other words, the effect of `fading memory' does not get captured, which can lead to incorrect estimation of deformation behaviour when the material relaxation time is comparable to the loading time-scale in the probing experiment. A typical way of accommodating this fading memory effect is to use the hereditary integral formulation with modulus functions as sum of a constant and a series of temporally exponentially decaying terms as the material constitutive model \cite{Ferry1980,Phan2017,Cho2016}, commonly called the Maxwell discrete relaxation spectrum or `Prony series' (PS). Often, only one constant and one exponentially decaying term suffice, and this special case is called as the `Standard Linear Solid' (SLS) model. Notably, the standard KV model may also be recovered from the SLS model, by using the Dirac-delta function.

The dimensional modulus functions, for the SLS model are\cite{Gutierrez2014,Huang2007}, 
\begin{subequations}
\label{eq:SLS_d}
\begin{equation}
\label{eq:K_SLS_d}
{K}(t^*) = {K}_E + {K}_{v}\exp\left(-{t^*}/{\tau_{k}}\right),
\end{equation}  
\begin{equation}
\label{eq:G_SLS_d}
{G}(t^*) = {G}_E + {G}_{v}\exp\left(-{t^*}/{\tau_{g}}\right),
\end{equation}  
\end{subequations}
and for the KV model are, 
\begin{subequations}
\label{eq:KV_d}
\begin{equation}
\label{eq:K_KV_d}
{K}(t^*) = {K}_E + {K}_{v}\tau_{k}\delta(t^*),
\end{equation}  
\begin{equation}
\label{eq:G_KV_d}
{G}(t^*) = {G}_E + {G}_{v}\tau_{g}\delta(t^*),
\end{equation}  
\end{subequations}

For both the KV and SLS models, we choose $\displaystyle {K}_c = {K}_E+\tau_{k}{K}_v$, $\displaystyle {G}_c = {G}_E+\tau_{g}{G}_v$.

Using equations \eqref{eq:SLS_d} and equation \eqref{eq:coeffs}, the Fourier space bulk and shear moduli for the SLS model are obtained as,
\begin{subequations}
\label{eq:bar_SLS_nd}
\begin{equation}
\label{eq:barK_SLS_nd}
\bar{\mathcal{K}}(f) = \frac{{K}_E + \frac{i2\pi\omega\tau_k f}{1+i2\pi\omega\tau_k f}{K}_v}{3K_c+4G_c} ,
\end{equation}  
\begin{equation}
\label{eq:barG_SLS_nd}
\bar{\mathcal{G}}(f) = \frac{{G}_E + \frac{i2\pi\omega\tau_g f}{1+i2\pi\omega\tau_g f}{G}_v}{3K_c+4G_c},
\end{equation}  
\end{subequations}
Using equations \eqref{eq:KV_d} and equation \eqref{eq:coeffs}, the Fourier space bulk and shear moduli for the KV model are obtained as,
\begin{subequations}
\label{eq:bar_KV_nd_1}
\begin{equation}
\label{eq:barK_KV_nd_1}
\bar{\mathcal{K}}(f) = \frac{{K}_E + i2\pi\omega\tau_k f{K}_v}{3K_c+4G_c},
\end{equation}  
\begin{equation}
\label{eq:barG_KV_nd_1}
\bar{\mathcal{G}}(f) = \frac{{G}_E + i2\pi\omega\tau_g f{G}_v}{3K_c+4G_c}.
\end{equation}  
\end{subequations}

It is desirable to accommodate substrate material compressibility in a theoretical framework for viscoelasticity characterization, something that has been missing in the current state of mathematical modeling for the methodology we are considering. It is also important to mention here that while the Poisson's ratio alone is sufficient to represent the compressibility of a linear-elastic material, a counterpart single material constant for viscoelastic materials is not currently available because of their fading memory effect \cite{Hilton2011,Hilton2017}. Nonetheless, researchers have often considered Poisson's ratio as an empirical metric for viscoelastic materials as well in the interest of quantifying their compressibility \cite{Tschoegl2002}. In line with such definitions, we identify two ratios of viscoelastic material constants which are indicative of compressibility of the material. The first ratio, $\displaystyle \nu_{\rm E} = \frac{3-2\frac{{G}_E}{{K}_E}}{6+2\frac{{G}_E}{{K}_E}}$, is indicative of the compressibility of a viscoelastic material after a `long’ time from the application of a load when the material assumes its static deformed configuration. The second ratio, $\displaystyle \nu_{\rm v} = \frac{3-2\frac{{G}_v}{{K}_v}}{6+2\frac{{G}_v}{{K}_v}}$ is reminiscent of time-scales compatible to rapid temporal variations in the loading conditions typical to early transients in the characterization experiments.

\subsection{Solution Methodology}\label{subsec:Soln}

Equations \eqref{eq:l_H0}, \eqref{eq:Re_eq_simp_Fourier}, and \eqref{eq:phd_bcs_main_Fourier} constitute the set of equations representing the complete mathematical problem. The system response, i.e. the substrate deformation, the hydrodynamic pressure, and the force between sphere and substrate, is sinusoidal oscillatory\cite{Guan2017,Leroy2011}. In mathematical terms, for the loading $\displaystyle h = H-\frac{r^2}{2}  = \alpha \cos(t)$, the deflection, pressure, and force response are sinusoidal: $l = l_0\cos(t+\theta_l)$,  $p = p_0\cos(t+\theta_p)$ and $F = F_0\cos(t+\theta_F)$. These cosine functions may be represented equivalently as real parts of an exponential function with imaginary exponent, so that:
\begin{equation}
\label{eq:complex}
\begin{split}
h(t) = &  \text{Re}[\alpha\exp(it)] \\
l(r,t) = \text{Re}[\grave{l}(r,t)] = & \text{Re}[l_c(r)\exp(it)], \\
p(r,t) = \text{Re}[\grave{p}(r,t)] = & \text{Re}[p_c(r)\exp(it)], \\
\end{split}
\end{equation}
In these expressions, the accent $\grave{~~}$ signifies that the accented variable is the ‘complex counterpart’, and the actual variable is the real component of this complex counterpart. The subscript $c$ signifies complex amplitude.

We obtain complex counterparts of equations \eqref{eq:l_H0}, \eqref{eq:Re_eq_simp_Fourier}, and \eqref{eq:phd_bcs_main_Fourier}, corresponding to $\grave{l}(r,t)$ and $ \grave{p}(r,t)$ as,
\begin{subequations}
\label{eq:counterpart}
\begin{equation}
\label{eq:Re_eq_simp_fourier_counterpart}
\left[\frac{1}{12r}\frac{\rm d}{{\rm d}r}\left\{r\left(1+\frac{r^2}{2}\right)^3 \frac{{\rm d}p_{c}}{{\rm d}r}\right\} -  \frac{\Gamma}{\alpha}il_c - i  \right]  \cdot \text{Fr}[\exp(it)] = 0,
\end{equation}
\begin{equation}
\label{eq:phd_bcs_fourier_counterpart}
\begin{split}
\left[\left.\frac{{\rm d} p_c}{{\rm d} r}\right|_{r=0}\right]\cdot \text{Fr}[\exp(it)] = 0, \\ \left[\left.p_c\right|_{r\rightarrow \infty}\right]\cdot \text{Fr}[\exp(it)] \rightarrow 0.
\end{split}
\end{equation}
\begin{equation}
\label{eq:l_H0_counterpart}
[\tilde{l_c}(x) - \breve{X}(x,f) \tilde{p_c}(x)]\cdot \text{Fr}[\exp(it)] = 0.
\end{equation}
\end{subequations}

In equation \eqref{eq:counterpart}, the Fourier transform of $\exp(it)$, written as $\text{Fr}[\exp(it)]$, known to be the Dirac-delta function about $\displaystyle \frac{1}{2\pi}$, is non-zero (an impulse) at $\displaystyle f = \frac{1}{2\pi}$ and zero otherwise. Thus, equations \eqref{eq:Re_eq_simp_fourier_counterpart} to \eqref{eq:l_H0_counterpart} are trivially satisfied when $\displaystyle f \ne \frac{1}{2\pi}$. For these equations to be satisfied at $\displaystyle f = \frac{1}{2\pi}$, the terms in square braces on the left of each equation should be zero, i.e.,
\begin{subequations}
\label{eq:counterpart_simp}
\begin{equation}
\label{eq:Re_eq_simp_fourier_counterpart_simp}
\frac{1}{12r}\frac{\rm d}{{\rm d}r}\left\{r\left(1+\frac{r^2}{2}\right)^3 \frac{{\rm d}p_{c}}{{\rm d}r}\right\} = i\left(\frac{\Gamma}{\alpha}l_c + 1\right),
\end{equation}
\begin{equation}
\label{eq:phd_bcs_fourier_counterpart_simp}
\left.\frac{{\rm d} p_c}{{\rm d} r}\right|_{r=0} = 0, \left.p_c\right|_{r\rightarrow \infty} \rightarrow 0.
\end{equation}
\begin{equation}
\label{eq:l_H0_counterpart_simp}
\tilde{l}_c(x) = \breve{X}\left(x,f=\frac{1}{2\pi}\right) \tilde{p}_c(x).
\end{equation}
\end{subequations}

Equation \eqref{eq:counterpart_simp} is discretized into a system of non-linear algebraic equations, and is numerically solved using the multi-variable Newton Raphson method. 

The dimensionless expression for the delayed force response in the complex plane, $F_c$, is obtained from $p_c$ as, 
\begin{equation}
\label{eq:Fnd}
F_c = \displaystyle \int_{0}^{\infty}2\pi p_c(r)rdr,
\end{equation}
whereas its dimensional version reads: $F_c^*=\alpha\mu\omega R^2F_c$. The storage and loss moduli, $G^{\prime}$ and $G^{\prime\prime}$ respectively, are recovered from $F_c^*$ as,
\begin{equation}
G = G^{\prime} + i G^{\prime\prime} = \frac{F_c^*}{h_0}.
\end{equation}

\subsection{Distinction Index}\label{subsec:heuristics}

Distinction index is a quantification of how distinct the variation of $G^{\prime}$ and $G^{\prime\prime}$ with $D$ will be for two materials under identical experiment conditions, i.e. under identical set of parameters $R$, $L$, $h_0$, $\omega$, $\mu$. Following the analysis presented in section S7 of the ESI, we obtain distinction index, $\mathcal{D}_{\rm (A,B)}$, between two materials, `A' and `B', as,
\begin{equation}
\label{eq:similarity}
\begin{split}
\mathcal{D}_{\rm (A,B)} = & \left|{\frac{C_{\rm A}-C_{\rm B}}{|C_{\rm A}|+|C_{\rm B}|}}\right|,~~\text{where}\\
C_{\rm i} = & \frac{\breve{X}_{\rm i}(f=\frac{1}{2\pi})}{3{K}_{c(\rm i)}+4{G}_{c(\rm i)}},~~\rm i = \rm A,B.
\end{split}
\end{equation}

We next consider the two limiting substrate thicknesses, namely, (i) thin and (ii) semi-infinite. These limits correspond to $\bar{\beta}\ll 1$ and $\bar{\beta}\gg 1$, respectively, yielding the corresponding limiting cases of $\breve{X}$ (using equation \eqref{eq:X}), and subsequently the limiting expressions for $C_{i}$,
\begin{subequations}
\label{eq:C_limits}
\begin{equation}
\label{eq:C_thin}
C_{\rm i}^{\text{thin}~~~~} = \left.\frac{1}{(3{K}_{c(\rm i)}+4{G}_{c(\rm i)})(3\bar{\mathcal{K}}_{\rm i}+4\bar{\mathcal{G}}_{\rm i})}\right|_{f=\frac{1}{2\pi}},
\end{equation}
\begin{equation}
\label{eq:C_semiinfinite}
C_{\rm i}^{\text{semi-inf}} = \left.\frac{3\bar{\mathcal{K}}_{\rm i}+4\bar{\mathcal{G}}_{\rm i}}{(3{K}_{c(\rm i)}+4{G}_{c(\rm i)})\bar{\mathcal{G}}_{\rm i}(3\bar{\mathcal{K}}_{\rm i}+\bar{\mathcal{G}}_{\rm i})}\right|_{f=\frac{1}{2\pi}}.
\end{equation}
\end{subequations}
Equation \eqref{eq:C_limits} enables us to recover the distinction index in the limits of thin and semi-infinite substrate thickness, $\mathcal{D}_{\rm (A,B)}^{\text{thin}}$ and $\mathcal{D}_{\rm (A,B)}^{\text{semi-inf}}$ respectively, which are two key variables that can provide insights regarding designing the geometric and loading conditions to precisely characterize viscoelastic materials, as will be discussed in section \ref{subsec:heuristics_result} ahead.

\begin{table}
\begin{center}
\def~{\hphantom{0}}
\begin{tabular}{cccccccc}
\hline
{Candidate}				&	Line type					&
${K}_E$					&	${K}_v\tau_{k}$				&
${G}_E$					&	${G}_v\tau_{g}$				&
$\nu_{\rm E}$			&	$\nu_{\rm v}$				\\[3pt]
\hline
I						&	$\includegraphics[width=0.015\linewidth]{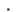}$	&
9.75 GPa 				& 	0.66 kPa-s 					&
195.0 kPa 				& 	13.30 mPa-s 				&
$\approx 0.5$			&	$\approx 0.5$				\\
II						&	$\includegraphics[width=0.025\linewidth]{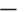}$	&
10.0 GPa			 	& 	2.05 kPa-s 					&
200.0 kPa			 	& 	40.93 mPa-s 				&
$\approx 0.5$			&	$\approx 0.5$				\\
III						&	$\includegraphics[width=0.025\linewidth]{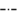}$	&
5.13 MPa 				& 	1.05 Pa-s 					&
208.0 kPa 				& 	42.56 mPa-s 				&
$0.48$					&	$0.48$						\\
IV						&	$\includegraphics[width=0.025\linewidth]{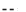}$	&
2.13 MPa 				& 	$~$435 mPa-s				&
220.0 kPa 				& 	45.02 mPa-s 				&
$0.45$					&	$0.45$						\\
V						&	$\includegraphics[width=0.025\linewidth]{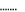}$	&
1.12 MPa 				& 	$~$229 mPa-s				&
240.0 kPa 				& 	49.11 mPa-s 				&
$0.4$					&	$0.4$						\\
\hline
\end{tabular}
\caption{Viscoelastic parameter values candidates for figure \ref{fig:G_KV} and associated discussion in section \ref{subsec:compress} and for figure \ref{subfig:similarity_KV_compress} and associated discussion in section \ref{subsec:heuristics}; all candidates are modeled as per the KV constitutive model; candidates I and II are the same as the first and second candidates in table 1 of Guan {\it et al} \cite{Guan2017}.}
\label{tab:compress}
\end{center}
\end{table}
\begin{table}
\begin{center}
\def~{\hphantom{0}}
\begin{tabular}{ccccc}
\hline
experiment 							&	Line \& Marker				&
$L$ 								&	$R$							&	
Thickness							\\[3pt]
condition							&	colour						&
(nm)								&	($\mu$m)					&	
regime								\\ 
\hline
A									&	$\includegraphics[width=0.010\linewidth]{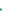}$ \textcolor{plotblue}{blue}			&
440									&	0.22						&	
semi-infinite ($L \gg \sqrt{DR}, ~ {\beta}/{\sqrt{\epsilon}} = 2.97 ~ { \rm to } ~ 9.38$)																					\\
B									&	$\includegraphics[width=0.010\linewidth]{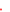}$ \textcolor{plotred}{red}			&
260									&	0.80						&	
thick ($L \sim \sqrt{DR}, ~ {\beta}/{\sqrt{\epsilon}} = 0.92 ~ { \rm to } ~ 2.91$)																					\\
C									&	$\includegraphics[width=0.010\linewidth]{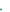}$ \textcolor{plotgreen}{green}		&
440									&	220						&	
thin ($L \ll \sqrt{DR}, ~ {\beta}/{\sqrt{\epsilon}} = 0.09 ~ { \rm to } ~ 0.3$)																					\\
\hline
\end{tabular}
\caption{Experiment conditions for figure \ref{fig:G_KV} and associated discussion in section \ref{subsec:compress}; $\omega$ = $2\pi \times 350 \times 10^3$ rad/s for each experiment; experiment conditions A and B are the same as those for the second and first candidates in table 1 of Guan {\it et al} \cite{Guan2017}.}
\label{tab:compress_loading}
\end{center}
\end{table}

\begin{figure}
\centering
\subfloat[\centering]{
{\includegraphics[width=0.95\linewidth]{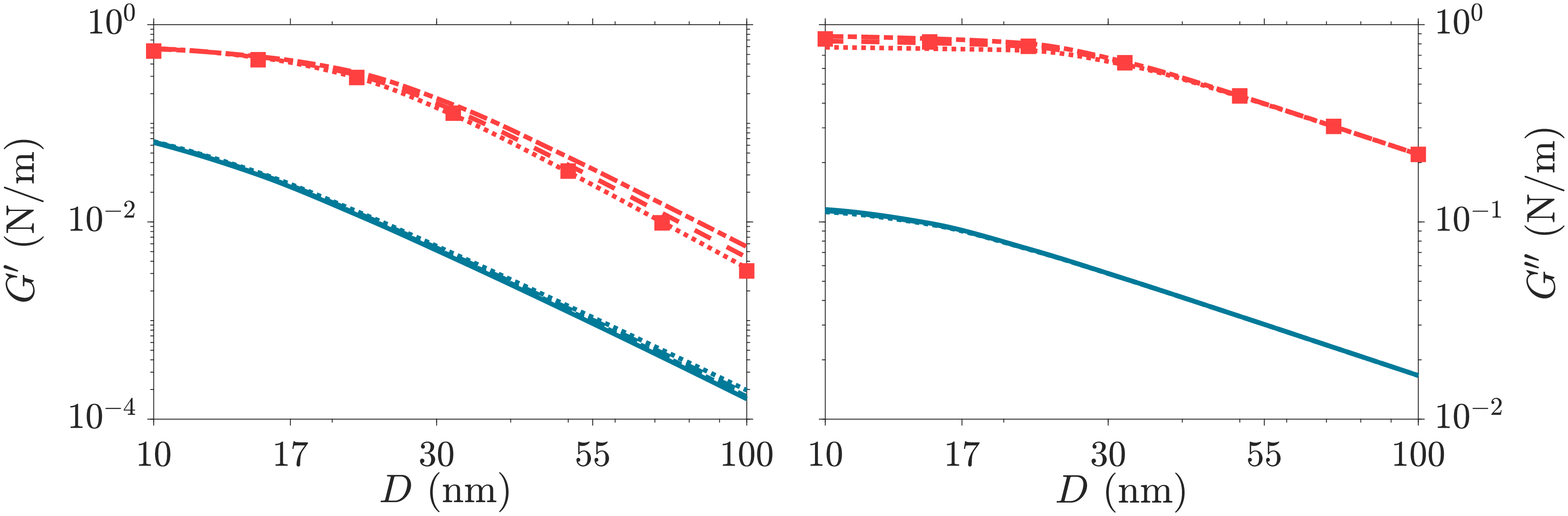}}
\label{subfig:G_Guan2017_same}
}\\
\subfloat[\centering]{
{\includegraphics[width=0.95\linewidth]{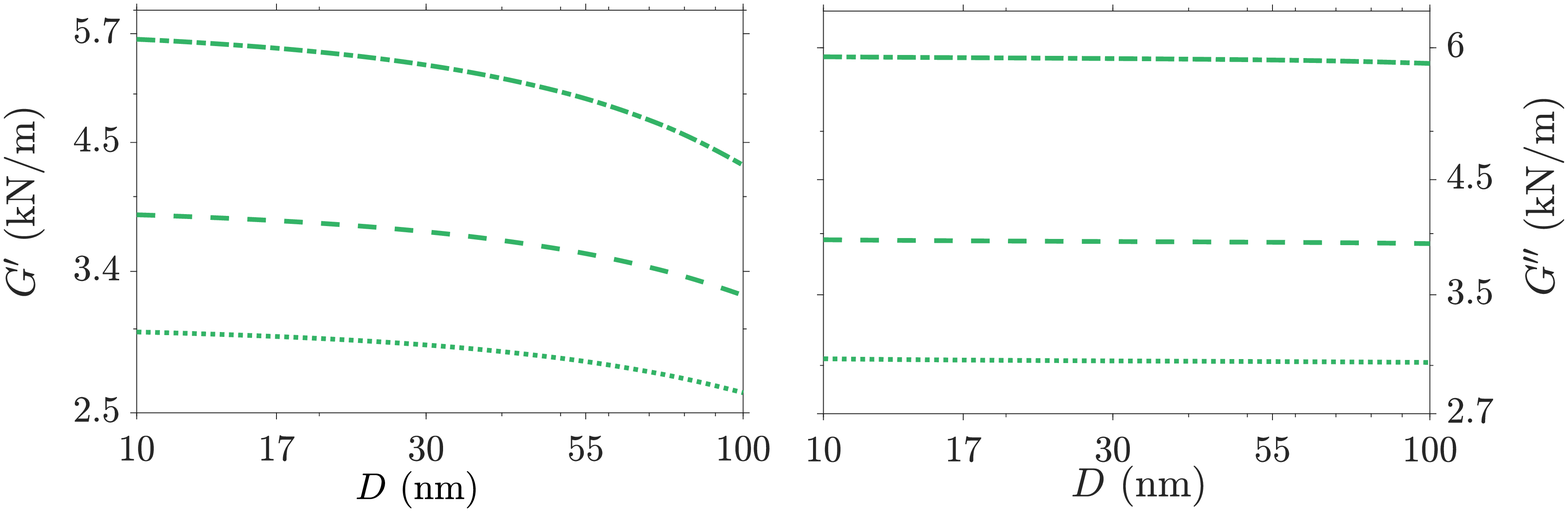}}
\label{subfig:G_Guan2017_distinct}
}
\caption{Variation of ${G}^{\prime}$ and ${G}^{\prime\prime}$ with $D$, for the viscoelastic parameter values candidates in table \ref{tab:compress} at the experiment conditions in table \ref{tab:compress_loading}; linetypes denote viscoelastic parameter values candidates (see table \ref{tab:compress} for details); line colors denote experiment conditions (see table \ref{tab:compress_loading} for details); all axes are log-scaled.}
\label{fig:G_KV}
\end{figure}

\section{Results and Discussion}\label{sec:results}

We demonstrate the implications of the present analytical framework towards drawing inferences from the experimental methodology of non-contact softness characterization of a substrate material using fluid mediated scanning probe microscopy (SPM)\cite{Leroy2011,Guan2017}. For this methodology, the key measurement is the variation of the force $F^*$ with decreasing magnitude of $D$. From this measurement, variations of $G^{\prime}$ and $G^{\prime\prime}$ with $D$, henceforth termed as `characteristics', are obtained and these serve as the primary characterization metric. These characteristics are dependent on the geometric and loading conditions on the one hand, and on the substrate viscoelasticity parameters on the other hand. The set of the parameters representing the geometric and loading conditions, $R$, $L$, $h_0$, and $\omega$ (termed as ‘experiment condition’ henceforth), is known \textit{a priori}. The set of substrate viscoelasticity parameters is unknown and has to be recovered using the experimentally obtained characteristics. To this end, one tallies the measured characteristics with the characteristics computed using a visco-elasto-hydrodynamic theory, to recover the constitutive model parameters. 

Characterization experiments at multiple loading conditions can be conducted for a particular material, and they are likely to yield different dynamic response characteristics. However, the recovered viscoelastic model parameter values should be consistent across the different characteristics, i.e. they should be invariant of the experimental condition. The foregoing analysis is specifically aimed to ensure this consistency via rationalizing the compatibility between physically appropriate modelling considerations and the data derived from controlled characterization experiments.

In section \ref{subsec:compress}, we first demonstrate an instance of possible inconsistency, or in other words, deviation from the above envisaged physically-consistent constitutive depiction. This pertains to a fallacious model prediction in which the KV viscosity of the substrate material varies unphysically with the substrate layer thickness by adopting previously reported theories. We subsequently show that by incorporating substrate material compressibility into the theoretical formalism as in the present model, we may rectify this anomaly. However, this corrective measure may adversely lead to a potential compromise in the predictive precision, which may be resolved by adhering to the `thin' regime of the substrate layer thickness.

On similar lines, in section \ref{subsec:relax}, we demonstrate another instance of possible inconsistency premised on previously reported theories – the fallacious dependence of the KV model parameters of the substrate material on the oscillation frequency of the surface probe. We subsequently show that by preferring the SLS model over the KV model, one may rectify this inconsistency. Our analysis reveals that the consequent potential compromise in precision can be resolved by appropriately tuning the oscillation frequency of the spherical probe. 

In section \ref{subsec:heuristics_result}, we demonstrate the utility of the distinction index in designing experiment conditions that will ensure supreme precision. 

For all the results presented and discussed in the subsequent subsections, we consider $\mu$ as 0.84 mPa and $h_0$ as 1 nm, to decipher the essential physics of interest.

\begin{table}
\begin{center}
\def~{\hphantom{0}}
\begin{tabular}{ccccccccc}
\hline
Candidate		&	Line 												&
${K}_E$			&	${K}_v\tau_{k}$			& 	$\tau_{k}$				&
${G}_E$			&	${G}_v\tau_{g}$			&	$\tau_{g}$				&
Constitutive															\\[3pt]
Candidate		&	type												&
(GPa)			&	(kPa-s)					& 	(ns)					&
(kPa)			&	(mPa-s)					&	(ns)					&
Model																	\\[3pt]
\hline
II				&	$\includegraphics[width=0.095\linewidth]{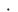}$	&
10.0		 	& 	2.05 	 					&	N/A												&
200.0		 	& 	40.93 		 				&	N/A												&	KV \\
VI				&	$\includegraphics[width=0.0285\linewidth]{demarcators_compress_candidate_I.eps}$		&
30.0 		 	& 	0.58 	 					&	N/A												&
600.0 		 	& 	11.58 		 				&	N/A												&	KV \\
VII				&	$\includegraphics[width=0.042\linewidth]{demarcators_compress_candidate_III.eps}$		&
10.0 		 	& 	2.05 	 					&	$~$72.4 										&
200.0 		 	& 	40.93 		 				&	$~$72.4 										&	SLS	\\
VIII			&	$\includegraphics[width=0.025\linewidth]{demarcators_compress_candidate_II.eps}$		&
10.0 		 	& 	2.05 	 					&	0.724 											&
200.0 		 	& 	40.93	 	 				&	0.724 											&	SLS \\
\hline
\end{tabular}
\caption{Viscoelastic parameter values candidates for section \ref{subsec:relax} and for figure \ref{subfig:similarity_KV_SLS} and associated discussion in section \ref{subsec:heuristics}; II and VI are modeled as per the KV constitutive model, and, VII and VIII are modeled as per the SLS constitutive model; $\nu_{\rm E} = \nu_{\rm v} \approx 0.5$ for each candidate.}
\label{tab:KV_SLS}
\end{center}
\end{table}
\begin{table}
\begin{center}
\def~{\hphantom{0}}
\begin{tabular}{ccc}
\hline
experiment							&	Line \& Marker 				&	$\omega$ 							\\
condition							&	colour						&	($\times 2\pi \times 10^3$ rad/s)	\\[3pt]
\hline
D									&	$\includegraphics[width=0.010\linewidth]{demarcators_compress_condition_A.eps}$ \textcolor{plotblue}{blue}		&			$0.1\times 350$			\\
E									&	$\includegraphics[width=0.010\linewidth]{demarcators_compress_condition_C.eps}$ \textcolor{plotgreen}{green}	&			$1\times 350$			\\
F									&	$\includegraphics[width=0.010\linewidth]{demarcators_compress_condition_B.eps}$ \textcolor{plotred}{red}		&			$10\times 350$			\\
\hline
\end{tabular}
\caption{Experiment conditions for the results of section \ref{subsec:relax}; $L$ = 440 nm, $R$ = 0.22 $\mu$m for each experiment conforming to the semi-infinite limit of substrate layer thickness ($L \gg \sqrt{DR}$).}
\label{tab:KV_SLS_loading}
\end{center}
\end{table}

\begin{figure}[!htb]
\centering
\subfloat[\centering]{
{\includegraphics[width=0.95\linewidth]{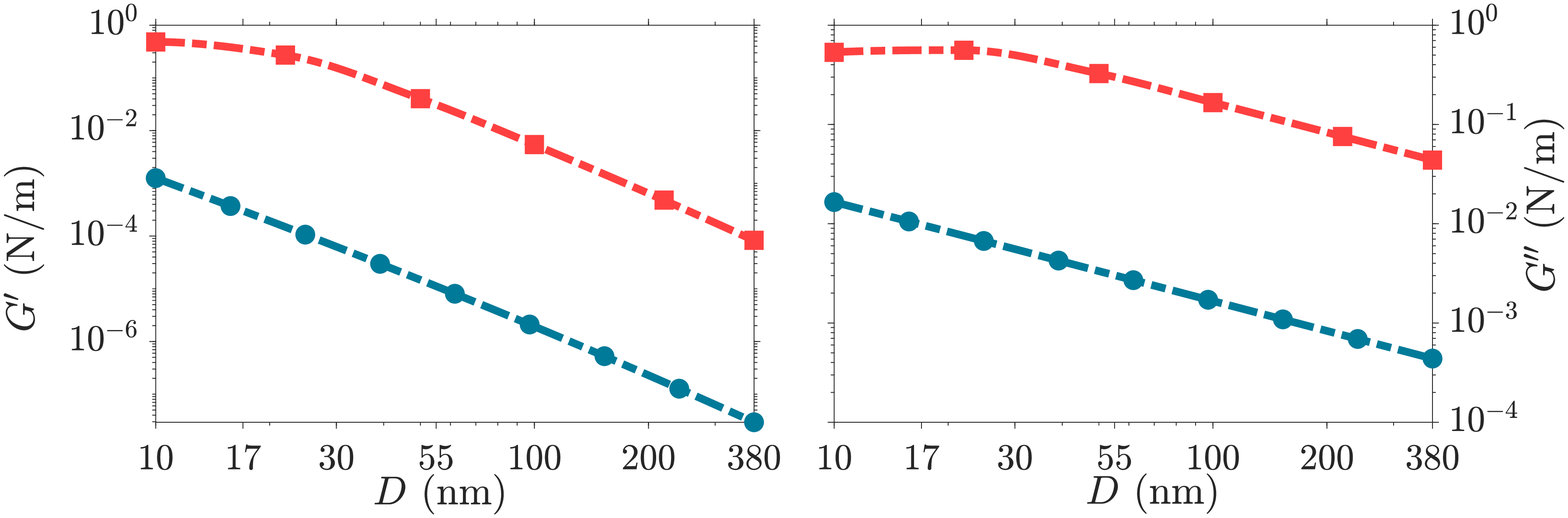}}
\label{subfig:G_KV_SLS}
}\\
\subfloat[\centering]{
{\includegraphics[width=0.95\linewidth]{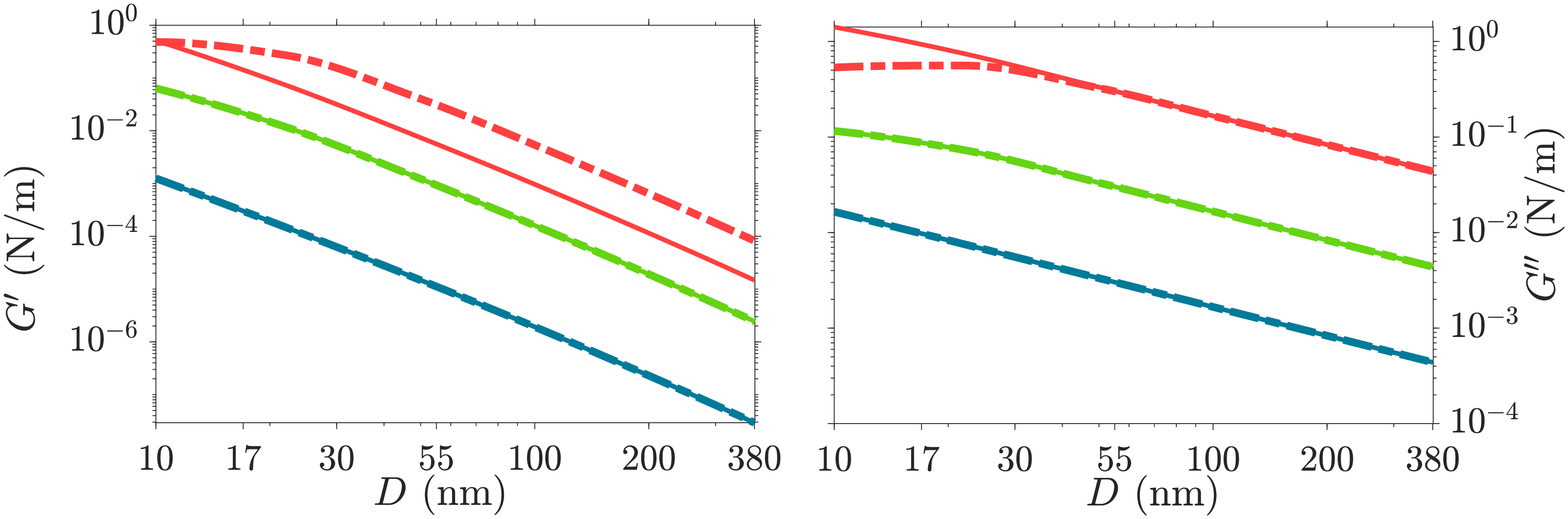}}
\label{subfig:G_KV_SLS_distinct}
}
\caption{Variation of ${G}^{\prime}$ and ${G}^{\prime\prime}$ with $D$, for the viscoelastic parameter values candidates in table \ref{tab:compress} at the experiment conditions in table \ref{tab:compress_loading}; linetypes denote viscoelastic parameter values candidates (see table \ref{tab:compress} for details); line colors denote experiment conditions (see table \ref{tab:compress_loading} for details); all axes are log-scaled.}
\label{fig:G_KV_SLS}
\end{figure}

\subsection{Effect of substrate material compressibility}\label{subsec:compress}

We first assess the system response characteristics for some key combinations of experimental conditions (table \ref{tab:compress_loading}) and candidates for viscoelastic model parameter values (table \ref{tab:compress}). The corresponding characteristics are presented in figure \ref{subfig:G_Guan2017_same}, which correspond to the experimental conditions A and B (see Table \ref{tab:compress_loading}). These are essentially the characteristics that were obtained by Guan {\it et al} \cite{Guan2017} from their experiments on a PDMS substrate. 

If we adhere to the consideration of incompressible substrate, the only plausible options are to map the material candidate I with the experimental condition A and the material candidate II with the experimental condition B. In other words, we are constrained to choose the KV model viscosity parameters to be variant with the experimental condition, for the same material. This inconsistency, as was encountered by Guan {\it et al} \cite{Guan2017}, however gets resolved with due accounting of the substrate compressibility as considered in the present model, opening up the possibilities of choosing the candidates III, IV, and V to consistently represent the experiment conditions A and B without necessitating any case-specific alteration in the KV model parameters. 

With the candidates III, IV and V open as possible options, there is a further scope of improvement based on the modelling considerations to pinpoint to that specific option which ensures the best possible precision in predicting the constitutive model parameters. Towards this, considering that the effect of substrate compressibility on its deformation is starker for thinner substrate layers Karan {\it et al} \cite{Karan2020a}, we may choose a set of values of $L$ and $R$ such that the substrate layer thickness falls under the thin regime, which corresponds to the experimental condition C. The resulting characteristics are presented in figure \ref{subfig:G_Guan2017_distinct}. Under the experimental condition C, if we obtain the characteristics represented by the dashed-dot line in this figure, the candidate III turns out to correspond to the precise set of viscoelastic parameter values for the substrate. Similarly, the candidate IV maps with the characteristics represented by the dashed line, and the candidate V for the dotted lines.

\subsection{Effect of probe oscillation frequency}\label{subsec:relax}

Another possible source of inconsistency in the viscoelastic model parametrization is an apparently fallacious dependence of the viscoelastic model parameters on the oscillation frequency of the probe. This may be eliminated by using the SLS instead of the KV model.

Towards establishing this proposition, we first assess the characteristics for some key combinations of the experimental conditions (table \ref{tab:KV_SLS_loading}) and candidates for the viscoelastic parameter values (table \ref{tab:KV_SLS}).For example, consider the dashed-dot lines in figure \ref{subfig:G_KV_SLS}, corresponding to the experimental conditions D and F. The KV model forces us to choose the candidate II for the experimental condition D and the candidate VI at experimental condition F, for the viscoelasticity model parametrization. A remedy that resolves this loading-condition-dependence is the consideration of the SLS model in place of the KV model, pinpointing the material-property candidate VII to be conforming to both the loading conditions D and F. However, the same is achieved by the candidate VIII as well, and such many-to-one mapping of the model parameters lacks quantitative specificity.

To distinguish between these two scenarios, we further observe that candidates VII and VIII differ in terms of the relaxation time. The sensitive dependence of the model behaviour on its relaxation time, however, turns out to be imperative only when the relaxation time is comparable to the time-scale of the imposed dynamics. The later time scale, which is inverse of the probe oscillation frequency, may be enhanced by tuning $\omega$, to arrive at the experiment condition F, with the resulting characteristics presented as the blue lines in figure \ref{subfig:G_KV_SLS_distinct}. Thus, conducting an experiment under the condition F, if we obtain the characteristics represented by the dashed-dot blue line, the candidate VII maps specifically to the corresponding material model parameters. On the other hand, for the solid blue line characteristics, the same maps to the candidate VIII.

\begin{figure}
\centering
\subfloat[\centering]{
{\includegraphics[width=0.45\linewidth]{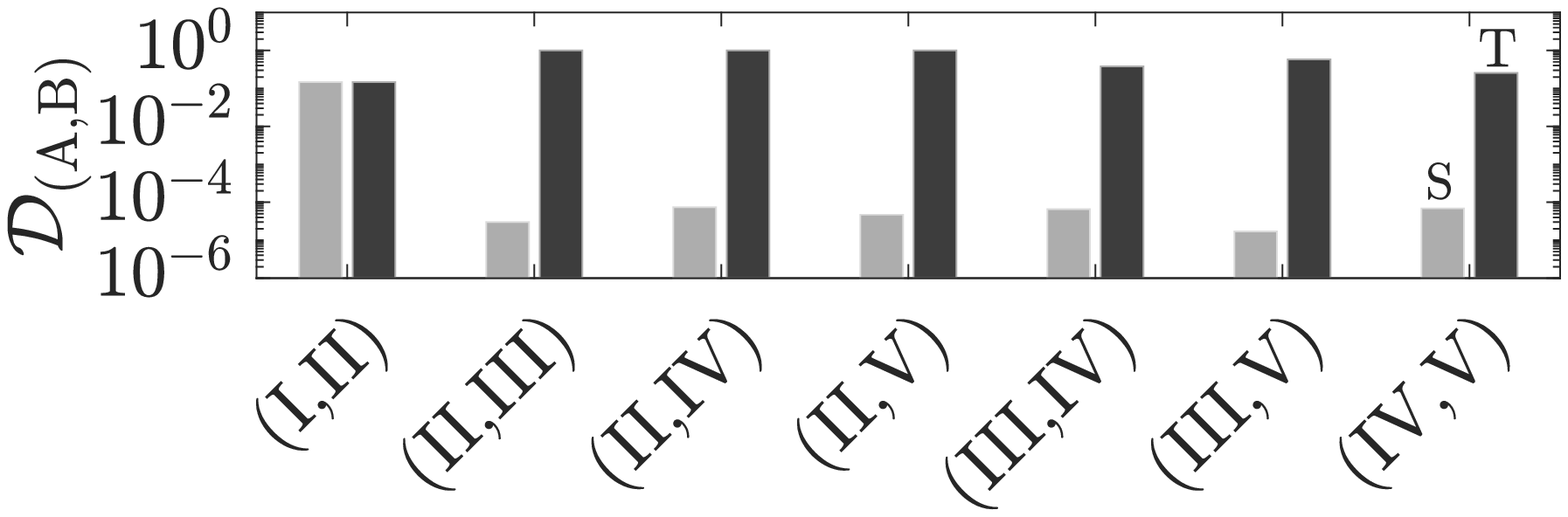}}
\label{subfig:similarity_KV_compress}
}
\subfloat[\centering]{
{\includegraphics[width=0.45\linewidth]{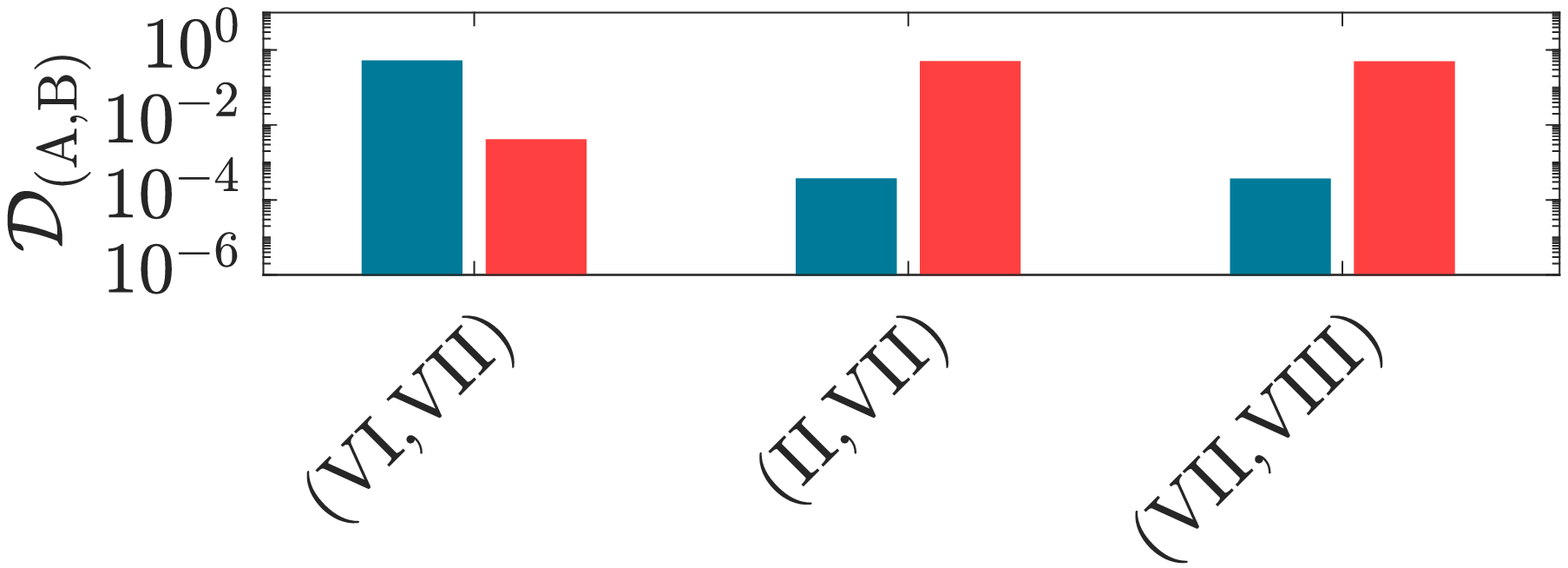}}
\label{subfig:similarity_KV_SLS}
}
\caption{Distinction indices for selected pairs of viscoelastic parameter values candidates corresponding to (a) section \ref{subsec:compress}, and, (b) section \ref{subsec:relax}; in panel a, the annotation `S' represents `semi-infinite' and the annotation `T' represents `thin', same order applying to all bar-pairs; the bar colors in panel b correspond to the color representation of experiment conditions in table \ref{tab:KV_SLS_loading}; for panel b, distinction index is identical for thin and semi-infinite limits; for each panel, the vertical axis is log-scaled.}
\label{fig:similarity}
\end{figure}

\subsection{Insights for designing experiment conditions}\label{subsec:heuristics_result}

In each of the sections \ref{subsec:compress} and \ref{subsec:relax}, we considered an assortment of characteristics for the viscoelastic parameter values, and observed that multiple material model candidates may yield identical characteristics for certain experimental conditions. We resorted to further considerations like preferential adherence to a regime of the substrate layer thickness or by tuning the oscillation frequency, to obviate such anomalous many-to-one mapping of material characterization parameters. However, while such considerations refer to physically consistent measures towards improving the specificity of the model parameter values, these remained to be rather subjective. 

For designing practical experiments, a more pinpointed quantitative parameter that ensures the envisaged specificity may be the distinction index as mentioned in section \ref{subsec:heuristics}. In figure \ref{subfig:similarity_KV_compress}, we present the distinction indices, for the semi-infinite (grey bars) and thin (black bars) regimes of the substrate layer thickness, for selected pairs of the candidates presented in table \ref{tab:compress_loading}. For a pair of incompressible material candidates, i.e. (I, II), the distinction indices can be seen identical for both the thin and semi-infinite regimes. However, for the rest of the pairs, having distinctive compressibility for the candidates in each pair, the distinction index for the thin regime is close to unity and is much higher than that for the semi-infinite regime. Thus, for a consistent quantitative accounting of the effect of substrate material compressibility, one needs to preferentially conduct the characterization experiments in the thin regime of the substrate layer thickness.

It is also imperative to recognize that a distinctive feature that renders the SLS model preferable over the KV model for the specific objectives of material characterization under fluid-mediated oscillatory loading is an accounting of the role of relaxation time of the substrate material and its interplay with the frequency of oscillation of the surface probe. This aspect is evidenced in figures \ref{subfig:similarity_KV_SLS}, where increasing the oscillation frequency leads to higher distinction index for the (II, VII) pair and decreasing the oscillation frequency leads to higher distinction index for the (VI, VII) pair.  We further observe that the two SLS candidates of table \ref{tab:KV_SLS} differ in terms of the relaxation time. Moreover, the distinction index for the (VII, VIII) pair increases as we increase the frequency from $2\pi\times 350\times 10^2$ rad/s to $2\pi\times 350\times 10^4$ rad/s. Upon close inspection, we observe that $\omega\tau_{g}\ll 1$ for both candidates VII and VIII at the lower  frequency but is $\sim 1$ for candidate VII and $\ll 1$ for candidate VIII at the higher frequency. Thus, to improve the precision in recovering the value of the substrate material relaxation time, one needs to tune the oscillation frequency of the surface probe such that the product $\omega\tau_{g}$ may act as a distinguishing parameter for the two potential material candidates.

\section{Conclusion}

We have presented a theoretical framework that enables the consistent and precise recovery of linear viscoelasticity parameters of a soft material as derived from contact free characterization experiments using surface probing apparatus. Our results demonstrate that due accounting for the substrate material compressibility as well as including the effect of the relaxation time in the description of the viscous modulus turn out to be two key interventions in eliminating inconsistent predictions from the relevant material characterization experiments. By pinpointing the implications of a decisive material characterization experiment, our results reveal the importance of exploring the thin substrate layer regime in addition to the usual thick substrate layer regime, as well as a tunable regime of the oscillation frequency, for depicting the material model parameters. These findings constitute the essential foundations of designing novel materials in multifarious applications ranging from engineering to biomedical technology.

% \backsection[Declaration of interests]{There are no conflicts to declare.}

% \backsection[Acknowledgements]{P.K., J.C., and S.C. acknowledge the financial support of the Ministry of Human Resource Development, Government of India. JC thanks the DST-INSPIRE program of the Government of India (DST/INSPIRE Faculty Award/2016/DST/ INSPIRE/04/2015/002825) as well as the ISIRD Funding (IIT/SRIC/ME/MFD/201718/70) from IIT Kharagpur for the computational resources. SC acknowledges Department of Science and Technology, Government of India, for Sir J. C. Bose National Fellowship.}

\appendix

\section{Flow Dynamics: Governing Equation and Boundary Conditions}\label{subsec:gdes}

The expanded form of the continuity equation (equation \eqref{eq:continuity_compact}) is,
\begin{equation}
\frac{1}{r^*} \frac{\partial (r^*v_{r}^*)}{\partial r^*} + \frac{\partial v_{z}^*}{\partial z^*} = 0,
\label{eq:ap_cont}
\end{equation}
and the Stokes equation (equation \eqref{eq:momentum_compact}) is,
\begin{subequations}
\begin{equation}
0 = \\ -\frac{\partial p^*}{\partial r^*} + \frac{\partial^2 v_{r}^*}{\partial z^{*2}} + \frac{1}{r^*} \frac{\partial}{\partial r^*} \left(r \frac{\partial v_{r}^*}{\partial r^*}\right) - \frac{v_{r}^*}{r^{*2}},
\label{eq:ap_rmom}
\end{equation}
\begin{equation}
0 = -\frac{\partial p^*}{\partial z^*} + \frac{\partial^2 v_{r}^*}{\partial z^{*2}} + \frac{1}{r^*} \frac{\partial}{\partial r^*} \left(r^* \frac{\partial v_{z}^*}{\partial r^*}\right).
\label{eq:ap_zmom}
\end{equation}
\label{eq:ap_mom}
\end{subequations}

\section{Substrate Deformation: Governing Equation and Boundary Conditions}\label{subsec:viscdef}

Upon expanding equation \eqref{eq:mecheq} using equation \eqref{eq:heriditary}, the two components of the mechanical equilibrium equation are obtained as,
\begin{subequations}
\label{eq:ap_disp_dim}
\begin{equation}
\label{eq:ap_rdisp_dim}
\begin{split}
\int_{-\infty}^{t^*} {\rm d}\tau^* \bigg[ & \left. 3{G}\frac{\partial }{\partial \tau^*}\left(\frac{\partial^2 u_{r}^*}{\partial \bar{z}^{*2}}\right) + \left(3{K}+{G}\right) \frac{\partial }{\partial \tau^*}\left(\frac{\partial^2 u_{\bar{z}}^*}{\partial r^* \partial \bar{z}^*}\right) + \right. \\ & \left. \left(3{K}+4{G}\right)  \frac{\partial }{\partial \tau^*}\left(\frac{\partial^2 u_{r}^*}{\partial r^{*2}} + \frac{1}{r^*} \frac{\partial u_{r}^*}{\partial r^*} - \frac{u_{r}^*}{r^{*2}} \right) \right]= 0,
\end{split}
\end{equation}
\begin{equation}
\begin{split}
\int_{-\infty}^{t^*} {\rm d}\tau^* \bigg[ & \left. \left(3{K}+4{G}\right)\frac{\partial }{\partial \tau^*}\left(\frac{\partial^2 u_{\bar{z}}^*}{\partial \bar{z}^{*2}}\right) +   \left(3{K}+{G}\right) \frac{\partial }{\partial \tau^*}\left(\frac{\partial^2 u_{r}^*}{\partial r^* \partial \bar{z}^*} + \frac{1}{r^*} \frac{\partial u_{r}^*}{\partial \bar{z}^*}\right) + \right. \\ & \left.  3{G}\frac{\partial }{\partial \tau^*}\left(\frac{\partial^2 u_{\bar{z}}^*}{\partial r^{*2}} + \frac{1}{r^*} \frac{\partial u_{\bar{z}}^*}{\partial r^*} \right) \right] = 0.
\end{split}
\label{eq:ap_ydisp_dim}
\end{equation}
\end{subequations}
Using the expression ${\underline{\underline{\sigma}}_{F}^*} =  -p^*{\underline{\underline{I}}}+\mu\left({\dot{\underline{\underline{E}}}_{F}^*}+{\dot{\underline{\underline{E}}}_{F}^{*{\text{T}}}}\right)$, the expanded form of traction balance condition (equation \eqref{eq:solid_conditions_traction_compact}) is,
\begin{subequations}
\begin{equation}
\begin{split}
\int_{-\infty}^{t^*} {\rm d}\tau^*  \bigg[ & \left. {G} \frac{\partial }{\partial \tau}\left(\frac{\partial u_{r}^*}{\partial \bar{z}^*} + \frac{\partial u_{\bar{z}}^*}{\partial r^*} \right)\right] =  \mu\left(\frac{\partial v_r^*}{\partial z^*}+\frac{\partial v_z^*}{\partial r^*} \right),
\end{split}
\label{eq:ap_rtrac_dim}
\end{equation} %%
\begin{equation}
\int_{-\infty}^{^*}  {\rm d}\tau^* \bigg[  \left. (3{K}+4{G})\frac{\partial}{\partial \tau^*}\left(\frac{\partial u_{\bar{z}}^*}{\partial \bar{z}^*}\right) +  (3{K}-2{G})\frac{\partial}{\partial \tau}\left(\frac{\partial u_{r}}{\partial r}+\frac{u_{r}}{r}\right) \right] =  -\left[p^*-2\mu\frac{\partial v_z^*}{\partial z^*}\right].
\label{eq:ap_ytrac_dim}
\end{equation}
\label{eq:ap_trac_dim}
\end{subequations}

\section{Scaling and Non-Dimensionalization}\label{subsec:scaling}

The governing equations and boundary conditions are non-dimensionalized with characteristic values of the system variables, as presented in table \ref{tab:nondim}. The non-dimensionalized variables, appearing henceforth, are represented by the same notation as their dimensional counterparts but with the superscript $^*$ dropped. Since we scale $H^*$ similar to $z^*$, the expression for $H$ is, 
\begin{equation}
\label{eq:H_nd_LSM}
H = 1+\frac{r^2}{2}+\alpha\cos(t).
\end{equation}
Since $\alpha \ll 1$, equation \eqref{eq:H_nd_LSM} gets further simplified to,
\begin{equation}
\label{eq:H_nd_LSM_simp}
H = 1+\frac{r^2}{2}.
\end{equation}

The scaling framework corresponding to the fluid domain is the classical scaling framework for lubrication squeeze flows; see Leal \cite{Leal2007}. 

The scaling of the different variables relevant to the substrate domain is consistent to the considerations outlined in Karan {\it et al} \cite{Karan2020a}, and the expressions for $\delta$ and $\theta$, as presented in the caption of table \ref{tab:nondim}, are obtained likewise.  

The non-dimensionalized governing equations and boundary conditions, following from the dimensional version in \ref{subsec:gdes}, are obtained as follows.

The continuity equation, \eqref{eq:ap_cont} retains its form,
\begin{equation}
\frac{1}{r} \frac{\partial (rv_{r})}{\partial r} + \frac{\partial v_{z}}{\partial z} = 0,
\label{eq:ap_cont_nd}
\end{equation}
The momentum-conservation equations, \eqref{eq:ap_mom}, transform into,
\begin{subequations}
\begin{equation}
% \frac{\epsilon^2\rho\omega R^2}{\mu}\left[\frac{\partial v_{r}}{\partial t} + \alpha\left(v_{r}\frac{\partial v_{r}}{\partial r}+v_{z}\frac{\partial v_{r}}{\partial z} \right)\right] 
0 =  -\frac{\partial p}{\partial r} + \frac{\partial^2 v_{r}}{\partial z^{2}} + \epsilon \left[ \frac{1}{r} \frac{\partial}{\partial r} \left(r \frac{\partial v_{r}}{\partial r}\right) - \frac{v_{r}}{r^{2}} \right],
\label{eq:ap_rmom_nd}
\end{equation}
\begin{equation}
% \frac{\epsilon^{3}\rho\omega R^{2}}{\mu}\left[k\frac{\partial v_{z}}{\partial t} + \alpha\left(v_{r}\frac{\partial v_{z}}{\partial r}+v_{z}\frac{\partial v_{z}}{\partial z} \right)\right] = 
0 = -\frac{\partial p}{\partial z} +  \epsilon\frac{\partial^2 v_{r}}{\partial z^{2}} + \epsilon^{2} \left[ \frac{1}{r} \frac{\partial}{\partial r} \left(r \frac{\partial v_{z}}{\partial r}\right) \right].
\label{eq:ap_zmom_nd}
\end{equation}
\label{eq:ap_mom_nd}
\end{subequations}

The corresponding boundary conditions get transformed to:
\begin{align}
v_r = 0, ~ v_z = -\sin(t) & \text{ \hspace{25pt} at \hspace{15pt} } z = H, \label{eq:bc_fluid_1_nd} \\
v_r = 0, ~ v_z = -\frac{\Gamma}{\alpha}\frac{\partial l}{\partial t} & \text{ \hspace{25pt} at \hspace{15pt} } z= -\Gamma l, \label{eq:bc_fluid_2_nd} \\
v_r\rightarrow 0,~v_z \rightarrow 0,~p_{\text{hd}} \rightarrow 0 & \text{ \hspace{25pt} as \hspace{15pt} } r \rightarrow \infty, \label{eq:bc_fluid_3_nd} \\
v_r = \frac{\partial v_z}{\partial r} = \frac{\partial p_{\text{hd}}}{\partial r} = 0 & \text{ \hspace{25pt} at \hspace{15pt} } r = 0, \label{eq:bc_fluid_4_nd}
\end{align}
with $\displaystyle \Gamma = \frac{\theta}{\epsilon}$.

To simplify the algebra corresponding the substrate domain, we normalize the modulus functions as: $\displaystyle \mathcal{K}(t-\tau) = \frac{{K}(t^*-\tau^*)}{3{K}_c+4{G}_c}$ and $\displaystyle \mathcal{G}(t-\tau) = \frac{{G}(t^*-\tau^*)}{3{K}_c+4{G}_c}$.

The two components of the mechanical equilibrium equation, equation \eqref{eq:ap_disp_dim}, thus transform into,
\begin{subequations}
\begin{equation}
\int_{-\infty}^{t} {\rm d}\tau  \bigg[ \left. \frac{\partial^2 u_{r}}{\partial \bar{z}^2} + \frac{\delta}{k^{\frac{1}{2}}\epsilon^{\frac{1}{2}}}\left(1+\frac{\lambda}{G}\right) \frac{\partial^2 u_{\bar{z}}}{\partial r \partial \bar{z}} + \frac{\delta^2}{k\epsilon} \left(2+\frac{\lambda}{G}\right) \left(\frac{\partial^2 u_{r}}{\partial r^{2}} + \frac{1}{r} \frac{\partial u_{r}}{\partial r} - \frac{u_{r}}{r^2} \right)\right] = 0,
\label{eq:ap_rdisp}
\end{equation}
\begin{equation}
\begin{split}
\int_{-\infty}^{t} {\rm d}\tau  \bigg[ & \left. \left(2+\frac{\lambda}{G}\right)\frac{\partial^2 u_{\bar{z}}}{\partial \bar{z}^2} + \frac{\delta}{k^{\frac{1}{2}}\epsilon^{\frac{1}{2}}}\left(1+\frac{\lambda}{G}\right) \left(\frac{\partial^2 u_{r}}{\partial r \partial \bar{z}} + \frac{1}{r} \frac{\partial u_{r}}{\partial \bar{z}}\right) + \right. \\ & \left. \frac{\delta^2}{k\epsilon} \left(\frac{\partial^2 u_{\bar{z}}}{\partial r^{2}} + \frac{1}{r} \frac{\partial u_{\bar{z}}}{\partial r} \right)\right] = 0.
\end{split}
\label{eq:ap_ydisp}
\end{equation}
\label{eq:ap_disp}
\end{subequations}

The corresponding boundary conditions get transformed to:
\begin{subequations}
\label{eq:bc_solid_nd_all}
\begin{equation}
\label{eq:bc_solid_1_nd}
u_r\rightarrow 0,~u_{\bar{z}} \rightarrow 0 \text{ \hspace{25pt} as \hspace{15pt} } r\rightarrow\infty,
\end{equation}
\begin{equation}
\label{eq:bc_solid_2_nd}
u_r = u_{\bar{z}} = 0 \text{ \hspace{25pt} at \hspace{15pt} } \bar{z} = \frac{\beta}{\delta},
\end{equation}
\begin{equation}
\label{eq:bc_solid_3_nd}
u_r = \frac{\partial u_{\bar{z}}}{\partial r} = 0 \text{ \hspace{25pt} as \hspace{15pt} } r=0,
\end{equation}
\end{subequations}
and the traction balance condition at the fluid-substrate interface,
\begin{subequations}
\begin{equation}
\int_{-\infty}^{t} {\rm d}\tau  \bigg[ \left. \mathcal{G} \frac{\partial }{\partial \tau}\left(\frac{\partial u_{r}}{\partial \bar{z}} + \frac{\delta}{\epsilon^{\frac{1}{2}}}\frac{\partial u_{\bar{z}}}{\partial r} \right)\right] = \frac{\delta\alpha}{\epsilon\theta}\frac{3\mu\omega}{(3{K}_c+4{G}_c)} \left[\left\{\frac{\epsilon^{\frac{1}{2}}}{3} \left(\frac{\partial v_r}{\partial z}+\epsilon\frac{\partial v_z}{\partial r}\right)  \right\}\right],
\label{eq:ap_rtrac_nd}
\end{equation}
\begin{multline}
\int_{-\infty}^{t} {\rm d}\tau \bigg[ \left. (3\mathcal{K}+4\mathcal{G})\frac{\partial}{\partial \tau}\left(\frac{\partial u_{\bar{z}}}{\partial \bar{z}}\right) + \right. \left.  \frac{\delta}{\epsilon^{\frac{1}{2}}}\cdot(3\mathcal{K}-2\mathcal{G})\frac{\partial}{\partial \tau}\left(\frac{\partial u_{r}}{\partial r}+\frac{u_{r}}{r}\right) \right] =  \\ -  \frac{\delta\alpha}{\epsilon\theta}\frac{3\mu\omega}{(3{K}_c+4{G}_c)}\left[p-\left\{2\epsilon\frac{\partial v_z}{\partial z}\right\}\right].
\label{eq:ap_ytrac_nd}
\end{multline}
\label{eq:ap_trac_nd}
\end{subequations}
We re-iterate here that the expression for $\theta$ is obtained by equating the scales of the fluid-side and substrate-side traction (i.e the left hand side and right hand side) of equation \eqref{eq:ap_ytrac_nd}.

\section{The Fluid Velocity Field}\label{subsec:simplifiedfluidgdes}

Owing to the assumptions 1 and 2 presented in section \ref{subsec:simp}, equation \eqref{eq:ap_mom_nd} simplifies to,
\begin{subequations}
\begin{equation}
0 = -\frac{\partial p}{\partial r} + \frac{\partial^2 v_{r}}{\partial z^{2}},
\label{eq:ap_rmom_nd_simp}
\end{equation}
\begin{equation}
0= -\frac{\partial p}{\partial z}.
\label{eq:ap_zmom_nd_simp}
\end{equation}
\label{eq:ap_mom_nd_simp}
\end{subequations}
Equations \eqref{eq:ap_cont_nd} and \eqref{eq:ap_mom_nd_simp} are solved following the traditional approach for lubrication problems. This leads to:
\begin{equation}
\label{eq:ap_vel_r}
v_r = \frac{1}{2}\frac{{\rm d} p}{{\rm d} r}\left(z^2-Hz\right).
\end{equation}
The above is substituted into equation \eqref{eq:ap_cont_nd} to get a first order ordinary differential equation for $v_z$. The same may be integrated to yield:
\begin{equation}
\label{eq:ap_vel_z}
\begin{split}
v_z =  & -\sin(t) - \frac{1}{12r}\frac{\partial }{\partial r}\left[r\frac{{\rm d} p}{{\rm d} r}\left\{2(z^3-H^3)- 3H (z^2-H^2)\right\}\right].
\end{split}
\end{equation}
Lastly, we apply equation \eqref{eq:bc_fluid_1_nd} to equation \eqref{eq:ap_vel_z} to get the Reynolds equation, equation \eqref{eq:Re_eq_simp}.

\section{Deflection-Pressure relation in Fourier-Hankel Space}\label{sec:ap_freqderiv}

Examining equations \eqref{eq:ap_disp} and \eqref{eq:ap_trac_nd}, we observe that the general expression,
\begin{equation}
\label{eq:general_hereditary}
\mathbb{M}(r,\bar{z},t) = \int_{-\infty}^{t}{\rm d}\tau\mathbb{G}(t-\tau)\frac{\partial \mathbb{W}(r,\bar{z},\tau)}{\partial \tau},
\end{equation}
is recurrent and represents all the terms in these equations with the appropriate forms of $\mathbb{G}(t-\tau)$ and $\mathbb{W}(r,\bar{z},\tau)$. Performing Fourier transformation, which is defined as,
\begin{equation}
\label{eq:fouriertransform}
\hat{\Phi}(f) =\text{Fr}[\Phi(t)] = \int_{-\infty}^{\infty}{\rm d}t\exp(-2\pi ift)\Phi(t),
\end{equation}
where $\hat{\Phi}(f)$ is the Fourier transform of the arbitrary function in $t$, $\Phi(t)$, and $f$ is the counterpart of $t$ in the Fourier space, on $\mathbb{M}(r,\bar{z},t)$ and employing principle of convolution for integral transformations, we arrive at the relation,
\begin{equation}
\label{eq:general_hereditary_Fourier_deriv}
\hat{\mathbb{M}}(r,\bar{z},f) = \bar{\mathbb{G}}(f)\hat{\mathbb{W}}(r,\bar{z},f),
\end{equation}
where $\displaystyle \bar{\mathbb{G}}(f) = 2\pi if\int_{0}^{\infty}{\rm d}t\exp(-2\pi ift)\mathbb{G}(t)$. The term $\mathbb{W}(r,z,t)$ stands in for the different expressions in equations \eqref{eq:ap_disp} and \eqref{eq:ap_trac_nd} which are of the form $\displaystyle \frac{\partial ^{m+n} u_{j}}{\partial r^m \partial \bar{z}^n}$, i.e., each expression consists of a single term which is some derivative of a displacement component. This approach of transforming the deformation behavior of a viscoelastic material from real space to Fourier-space follows the Elastic-Viscoelastic Correspondence Principle (EVCP) \cite{Phan2017,Tschoegl2002}. Out of this operation, we recover the Fourier space versions of equations \eqref{eq:ap_disp} and \eqref{eq:ap_trac_nd} as,
\begin{subequations}
\begin{equation}
3\bar{\mathcal{G}}\frac{\partial^2 \hat{u}_{r}}{\partial \bar{z}^2} + \left(3\bar{\mathcal{K}}+\bar{\mathcal{G}}\right) \frac{\delta}{\epsilon^{\frac{1}{2}}}\frac{\partial^2 \hat{u}_{\bar{z}}}{\partial r \partial \bar{z}} + \left(3\bar{\mathcal{K}}+4\bar{\mathcal{G}}\right) \frac{\delta^2}{\epsilon} \left(\frac{\partial^2 \hat{u}_{r}}{\partial r^{2}} + \frac{1}{r} \frac{\partial \hat{u}_{r}}{\partial r} - \frac{\hat{u}_{r}}{r^2} \right) = 0,
\label{eq:ap_rdisp_fourier}
\end{equation}
\begin{equation}
\left(3\bar{\mathcal{K}}+4\bar{\mathcal{G}}\right)\frac{\partial^2 \hat{u}_{\bar{z}}}{\partial \bar{z}^2} + \left(3\bar{\mathcal{K}}+\bar{\mathcal{G}}\right)  \frac{\delta}{\epsilon^{\frac{1}{2}}}\left(\frac{\partial^2 \hat{u}_{r}}{\partial r \partial \bar{z}} + \frac{1}{r} \frac{\partial \hat{u}_{r}}{\partial \bar{z}}\right) + 3\bar{\mathcal{G}}\frac{\delta^2}{\epsilon} \left(\frac{\partial^2 \hat{u}_{\bar{z}}}{\partial r^{2}} + \frac{1}{r} \frac{\partial \hat{u}_{\bar{z}}}{\partial r} \right) = 0.
\label{eq:ap_ydisp_fourier}
\end{equation}
\label{eq:ap_disp_fourier}
\end{subequations}
\begin{subequations}
\begin{equation}
\bar{\mathcal{G}} \left(\frac{\partial \hat{u}_{r}}{\partial \bar{z}} + \frac{\delta}{\epsilon^{\frac{1}{2}}}\frac{\partial \hat{u}_{\bar{z}}}{\partial r} \right) = 0,
\label{eq:ap_rtrac_simp_fourier}
\end{equation}
\begin{equation}
\frac{\partial \hat{u}_{\bar{z}}}{\partial \bar{z}} +  \left(\frac{3\bar{\mathcal{K}}-2\bar{\mathcal{G}}}{3\bar{\mathcal{K}}+4\bar{\mathcal{G}}}\right)\frac{\delta}{\epsilon^{\frac{1}{2}}}\left(\frac{\partial \hat{u}_{r}}{\partial r}+\frac{\hat{u}_{r}}{r}\right) = -\frac{\hat{p}}{(3\bar{\mathcal{K}}+4\bar{\mathcal{G}})},
\label{eq:ap_ytrac_simp_fourier}
\end{equation}
\label{eq:ap_trac_simp_fourier}
\end{subequations}
where,
\begin{equation}
\begin{split}
\bar{\mathcal{K}}(f) = 2\pi if\int_{0}^{\infty}{\rm d}t\exp(-2\pi ift)\mathcal{K}(t), \\
\bar{\mathcal{G}}(f) = 2\pi if\int_{0}^{\infty}{\rm d}t\exp(-2\pi ift)\mathcal{G}(t).
\end{split}
\label{eq:bar_moduli}
\end{equation}
The terms in curly braces in equation \eqref{eq:ap_trac_nd} have been dropped on the basis of assumption 2 presented in section \ref{subsec:simp}.

After performing Fourier transformation on equation \eqref{eq:bc_solid_nd_all}, we collect the same with equations \eqref{eq:ap_disp_fourier} and \eqref{eq:ap_trac_simp_fourier}. Following Hankel-space analysis similar to Karan {\it et al} \cite{Karan2020a}, we obtain the relation between deflection and pressure in the Fourier-Hankel space, \begin{equation}
\label{eq:l_H0_1}
\tilde{\hat{l}} = \breve{X}\tilde{\hat{p}}.
\end{equation}
The above is presented as equation \eqref{eq:l_H0} in the main text. Here, $~\tilde{}~$ represents zeroeth-order Hankel transformation,
\begin{equation}
\label{eq:Hankel}
\tilde{\Psi}(x) = H_{0}[\Psi(r)] = \int_{0}^{\infty}{\rm d}r\cdot rJ_{0}(xr)\cdot \Psi(r).
\end{equation}
$\tilde{\Psi}(x)$ is the zeroeth-order Hankel transform of the arbitrary function in $r$, $\Psi(r)$, and $x$ is the counterpart of $r$ in the Hankel space. 

\section{Similarity Index}\label{sec:similarity}

If the variation of $G^{\prime}$ and $G^{\prime\prime}$ with $D$ are to be identical for two arbitrary materials A and B under a specific set of geometric and loading conditions, i.e. the set of parameters $L$, $R$, $h_0$, $\omega$, $\mu$, then,
\begin{enumerate}
\item $F_c^*$ should be identical for the two materials.
\item As a consequence of (1) above, $\displaystyle F_c = \frac{F_c^*}{\mu\omega\alpha R^2}$ should be identical for the two materials.
\item Since $\displaystyle F_c = \int_{0}^{\infty}2\pi p_c r dr$, a sufficient condition for (2) as above to be satisfied is that $p_c$ is identical for the two materials.
\item Taking a look at Reynolds equation for $p_c$, equation \eqref{eq:Re_eq_simp_fourier_counterpart_simp}, we observe that the above leads to:
\begin{equation}
\label{eq:similarity_1}
\begin{split}
\left(\frac{\Gamma}{\alpha}l_c+1\right)_{\rm A} & = \left(\frac{\Gamma}{\alpha}l_c+1\right)_{\rm B} \\
\implies \Gamma_{\rm A} l_{c \rm (A)}+\alpha & = \Gamma_{\rm B}l_{c \rm (B)}+\alpha \\
\implies \Gamma_{\rm A} l_{c \rm (A)} & = \Gamma_{\rm B}l_{c \rm (B)} \\
\implies \theta_{\rm A} l_{c \rm (A)} & = \theta_{\rm B}l_{c \rm (B)} \\
\end{split}
\end{equation}
\item 
Equivalences of $p_c$ and $\tilde{p}_c$ as well as $l_c$ and $\tilde{l}_c$ are notable.
\item The expression for $l_c$, in Hankel space, is,
\begin{equation}
\label{eq:similarity_2}
\begin{split}
\tilde{l}_c & = \breve{X}\left(f=\frac{1}{2\pi}\right)\tilde{p}_c \\
\implies \theta \tilde{l}_c & = \theta \breve{X}\left(f=\frac{1}{2\pi}\right)\tilde{p}_c \\
\end{split}
\end{equation}
\item Applying the above for the two materials, we get:
\begin{equation}
\label{eq:similarity_3}
\begin{split}
\theta_{\rm A}\breve{X}_{\rm A}\left(f=\frac{1}{2\pi}\right) & = \theta_{\rm B}\breve{X}_{\rm B}\left(f=\frac{1}{2\pi}\right) \\
\implies \frac{\breve{X}_{\rm A}\left(f=\frac{1}{2\pi}\right)}{3{K}_{c \rm (A)}+4{G}_{c \rm (A)}} & = \frac{\breve{X}_{\rm B}\left(f=\frac{1}{2\pi}\right)}{3{K}_{c \rm (B)}+4{G}_{c \rm (B)}} \\
\implies C_{\rm A} & = C_{\rm B}, \\
{\rm where } &~~~C = \frac{\breve{X}\left(f=\frac{1}{2\pi}\right)}{3{K}_{c}+4{G}_{c}}
\end{split}
\end{equation}
\end{enumerate}

\bibliographystyle{asmems4}
\bibliography{refs_short}

\end{document}